\documentclass[12pt]{article}
\usepackage[english]{babel}
\usepackage{indentfirst}
\usepackage{amsmath,amssymb,amsthm,amsfonts,amscd}
\usepackage{graphicx}

\usepackage{amssymb}

\begin{document}

\newcommand{\beqn}{\begin{eqnarray}}
 \newcommand{\eeqn}{\end{eqnarray}}
 \newcommand{\be}{\begin{equation}}
 \newcommand{\ee}{\end{equation}}
 \newcommand{\ba}{\begin{array}}
 \newcommand{\ea}{\end{array}}

 \newcommand{\ve}{\varepsilon}
\newcommand{\E}{\mathbb{E}}
\newcommand{\T}{\mathbb{T}}
\newcommand{\R}{\mathbb{R}}
\newcommand{\Z}{\mathbb{Z}}
\newcommand{\N}{\mathbb{N}}

\newcommand{\pa}{\partial}
\newcommand{\fr}{\frac}
\newcommand{\ds}{\displaystyle}
\newcommand{\const}{\mathop{\rm const}\nolimits}
\newcommand{\supp}{\mathop{\rm supp}\nolimits}

 \renewcommand{\theequation}{\thesection.\arabic{equation}}
 \newtheorem{theorem}{Theorem}[section]
 \newtheorem{definition}[theorem]{Definition}
  \newtheorem{lemma}[theorem]{Lemma}
 \newtheorem{example}[theorem]{Example}
 \newtheorem{remark}[theorem]{Remark}
 \newtheorem{remarks}[theorem]{Remarks}
 \newtheorem{cor}[theorem]{Corollary}
 \newtheorem{pro}[theorem]{Proposition}

\newcommand{\bo}{{\hfill\loota}}
\newcommand{\loota}{\hbox{\enspace{\vrule height 7pt depth 0pt width 7pt}}}

 \begin{center}
{\Large\bf
 Large-time asymptotic behavior of  the infinite system  of harmonic oscillators
 on the half-line}
\vspace{1cm}\\
{\large T.V. Dudnikova}
\footnote{
The work was supported partly by the research grants of
Russian Science Fondation (Grant No. 14-21-00025)  and of RFBR (Grant No. 15-01-03587)
}\medskip\\
{\it Keldysh institute of Applied Mathematics RAS\\
Miuskaya sq. 4, Moscow 125047, Russia}\medskip\\
E-mail:~tdudnikov@mail.ru
\end{center}
\vspace{1cm}

\begin{abstract}
The mixed initial-boundary value problem for infinite
one-dimensional chain of harmonic oscillators on the half-line is considered.
We study the large time  behavior of solutions
and derive the dispersive bounds.
\medskip

\end{abstract}


\section{Introduction}
We consider the infinite system of harmonic oscillators
on the half--line:
\beqn
\ddot u(x,t)=(\nu^2\Delta_L-m^2) u(x,t),\quad x\in\N,\quad t>0,
\label{1.1}
\eeqn
with the boundary condition (as $x=0$)
\beqn\label{1.2}
\ddot u(0,t)=\nu^2(u(1,t)-u(0,t))-m^2u(0,t)-\kappa u(0,t)-\gamma\dot u(0,t),
\quad t>0,
\eeqn
and with the initial condition (as $t=0$)
\beqn\label{1.3}
u(x,0)=u_0(x),\quad \dot u(x,0)=v_0(x),\quad x\ge0.
\eeqn
Here $u(x,t)\in\R$, $\nu>0$, $m,\kappa,\gamma\ge0$,
 $\Delta_L$ denotes the second derivative on $\Z$:
$$
\Delta_L u(x)=u(x+1)-2u(x)+u(x-1),\quad x\in\Z.
$$
If $\gamma=0$, then formally the system (\ref{1.1})--(\ref{1.2}) is
Hamiltonian with  the Hamiltonian functional
 \beqn\label{H}
 {\rm H}(u,\dot u):=\ds\frac{1}{2}
\sum\limits_{x\ge0}\Big(
|\dot u(x,t)|^2+\nu^2|u(x+1,t)-u(x,t)|^2+m^2|u(x,t)|^2\Big)+\frac12\kappa|u(0,t)|^2.
\eeqn
We assume that the initial data $Y_0(x)=(u_0(x),v_0(x))$ belong to
the Hilbert space ${\cal H}_{\alpha,+}$, $\alpha\in\R$, defined below.
 \begin{definition} \label{d1.1'}
(i) $\ell^2_{\alpha,+}\equiv\ell^2_{\alpha,+}(\Z_+)$,
$\alpha\in\R$, is the  Hilbert space of sequences $u(x)$, $x\ge0$,
 with  norm
$ \Vert u\Vert^2_{\alpha,+}
=\sum\limits_{x\ge0}|u(x)|^2\langle x\rangle^{2\alpha}<\infty$,
$\langle x\rangle:=(1+x^2)^{1/2}$.\\
(ii)
$ {\cal H}_{\alpha,+}=\ell^2_{\alpha,+}\otimes\ell^2_{\alpha,+}$
is the  Hilbert space of pairs $Y=(u,v)$ of sequences equipped with  norm
 $ \Vert Y\Vert^2_{\alpha,+}
= \Vert u\Vert^2_{\alpha,+}+ \Vert v\Vert^2_{\alpha,+}<\infty$.
 \end{definition}

On the coefficients $m,\kappa,\nu,\gamma$ of the system
we impose condition {\bf C} or ${\bf C}_0$.
\begin{description}
\item[${\bf C}$]
If $\gamma\not=0$,  then  $m$ or $\kappa$ is not zero.

  In addition, if $\gamma\in(0,\nu)$ and $m=0$, then $\kappa\not=2(\nu^2-\gamma^2)$;

 if $\gamma\in\left(0,\left(\sqrt{m^2+4\nu^2}-m\right)/2\right]$ and $m\not=0$,
 then $\kappa\not=\nu^2-\gamma^2\pm\sqrt{(\nu^2-\gamma^2)^2-m^2\gamma^2}$.

 If $\gamma=0$, then $\kappa\in(0,2\nu^2)$.

 \item[${\bf C}_0$]
  $\gamma=0$ and $\kappa=2\nu^2$ or
  $\gamma=\kappa=0$ and  $m\not=0$.
   \end{description}

The main objective of the paper is to prove  that
 for any initial data $Y_0\in{\cal H}_{\alpha,+}$ with $\alpha>3/2$,
  the solution $Y(t)=(u(\cdot,t),\dot u(\cdot,t))$ of the system obeys the following bound
\be\label{0.3}
\Vert Y(t)\Vert_{-\alpha,+}\le C\langle t\rangle^{-\beta/2}\Vert Y_0\Vert_{\alpha,+},
\quad t\in\mathbb{R},
\ee
where $\beta=3$ if  {\bf C} holds, and $\beta=1$ if
 ${\bf C}_0$ holds.
We specify the  behavior of the solutions as $t\to\infty$ in
Theorem~\ref{thC}.
\smallskip

For the solutions of the linear
discrete Schrodinger and Klein--Gordon equations in the whole space,
the dispersive estimates of the type (\ref{0.3}) were obtained  by Shaban and Vainberg \cite{ShV},
 Komech, Kopylova and Kunze \cite{KKK} and  Pelinosky and Stefanov \cite{PS}.
The wave operators for the discrete Schrodinger operators were studied by Cuccagna \cite{C}.
In \cite{D16}, we considered the linear Hamiltonian  system consisting of
the  discrete Klein--Gordon field  coupled to a particle
and  obtained the similar results on the long--time behavior  for the solutions.
 In \cite{jmp-2017}, the considered model (\ref{1.1})--(\ref{1.3})
was studied with {\em random} initial data $Y_0\in {\cal H}_{\alpha,+}$ with $\alpha<-3/2$.
In this paper, the model is studied with initial data from the space
${\cal H}_{\alpha,+}$ with $\alpha>3/2$, and
the long time asymptotics of the solutions are constructed.

\setcounter{equation}{0}
\section{Main Results}\label{sec2}
The existence and uniqueness of the solutions to the problem
 (\ref{1.1})--(\ref{1.3}) was proved in \cite{jmp-2017}.
\begin{theorem}\label{T.A}
Let $\gamma,\kappa,m\ge0$, $\nu>0$,
 and let $Y_0\in{\cal H}_{\alpha,+}$, $\alpha\in\R$.
Then the problem (\ref{1.1})--(\ref{1.3}) has
a unique solution $Y(t)\in C(\R,{\cal H}_{\alpha,+})$. The operator
$U(t):Y_0\to Y(t)$ is continuous on ${\cal H}_{\alpha,+}$. Moreover,
 there exist constants $C,B<\infty$
such that $\Vert U(t)Y_0\Vert_{\alpha,+}\le Ce^{B|t|}\Vert Y_0\Vert_{\alpha,+}$, $t\in\R$.
For  $Y_0\in{\cal H}_{0,+}$, the following bound holds,
\be\label{H-1}
H(Y(t))+\gamma\int_0^t|\dot u(0,s)|^2\,ds=H(Y_0),\quad t\in\R,
\ee
where $H(Y(t))$ is defined in (\ref{H}).
\end{theorem}

The proof is based on the following representation for
 the solution $u(x,t)$ of the problem (\ref{1.1})--(\ref{1.3}):
\be\label{2.1}
u(x,t)=z(x,t)+q(x,t),\quad x\ge0,\quad t>0,
\ee
where $z(x,t)$  is a solution of the mixed problem
with zero boundary condition,
\beqn
&&\ddot z(x,t)=(\nu^2\Delta_L-m^2)z(x,t),\quad x\in\N,\quad t>0,
\label{a.1}\\
&&z(0,t)=0,\quad t\ge0,
\label{a.2}\\
&&z(x,0)=u_0(x),\quad \dot z(x,0)=v_0(x),\quad x\in\N.
\label{a.3}
\eeqn
 Therefore, $q(x,t)$  is a solution of the following mixed problem
\beqn
&&\ddot q(x,t)=(\nu^2\Delta_L-m^2)q(x,t),\quad x\in\N,\quad t>0,\label{b.1}\\
&&\ddot q(0,t)=\nu^2(q(1,t)-q(0,t))-(m^2+\kappa) q(0,t)-\gamma\dot q(0,t) +
\nu^2 z(1,t),\,\,\,\,\label{b.2}\\
&&q(x,0)=0,\quad \dot q(x,0)=0,\quad x\in\N, \label{b.3}\\
&&q(0,0)=u_0(0),\quad \dot q(0,0)=v_0(0). \label{b.4}
\eeqn

We state the results concerning the solutions of
the problem (\ref{a.1})--(\ref{a.3}).
\begin{lemma} \label{l2.1} (see Lemma 2.7 in \cite{D08})
Assume that $\alpha\in\R$. Then
 for any  $Y_0 \in {\cal H}_{\alpha,+}$, there exists  a unique solution
$Z(t)\equiv(z(\cdot,t),\dot z(\cdot,t))\in C(\R, {\cal H}_{\alpha,+})$  to the mixed problem
(\ref{a.1})--(\ref{a.3});
the operator  $U_0(t):Y_0\mapsto Z(t)$ is continuous
on ${\cal H}_{\alpha,+}$.
Furthermore, the following bound holds,
\be\label{c.2}
\Vert U_0(t) Y_0\Vert_{\alpha,+}\le C\langle t\rangle^\sigma\Vert Y_0\Vert_{\alpha,+},
\ee
with some constants $C=C(\alpha),\sigma=\sigma(\alpha)<\infty$.
\end{lemma}

The proof of Lemma~\ref{l2.1} is based on the following formula for
the solution  $Z(x,t)=(Z^0(x,t),Z^1(x,t))\equiv(z(x,t),\dot z(x,t))$
 of the problem (\ref{a.1})--(\ref{a.3}):
\be\label{sol}
Z^i(x,t)=\sum\limits_{j=0,1}\sum\limits_{x'\ge1}
{\cal G}^{ij}_{t,+}(x,x') Y_0^j(x'), \quad x\in \Z_+,
\ee
where
the Green function ${\cal G}_{t,+}(x,x')=({\cal G}_{t,+}^{ij}(x,x'))_{i,j=0}^1$ is
\beqn\label{3.2}
 {\cal G}^{ij}_{t,+}(x,x'):={\cal G}^{ij}_t(x-x')-{\cal G}^{ij}_t(x+x'),
\quad
{\cal G}^{ij}_t(x)\equiv\frac1{2\pi}\int_{\T}
e^{-ix \theta} \hat{\cal G}^{ij}_t(\theta)\,d\theta,
\eeqn
\beqn\label{hatcalG}
(\hat{\cal G}_t^{ij}(\theta))_{i,j=0}^1
=\left(\ba{ll} \cos\phi(\theta)t&
\frac{\sin\phi(\theta)t}{\phi(\theta)}\\
-\phi(\theta)\sin\phi(\theta)t& \cos\phi(\theta)t
\ea\right),\,\,\, \phi(\theta)=\sqrt{\nu^2(2-2\cos\theta)+m^2}.
\eeqn
In particular,
$\phi(\theta)=2\nu|\sin(\theta/2)|$ if $m=0$.
We see that $Z(0,t)\equiv 0$ for any $t$, since
${\cal G}^{ij}_t(-x)={\cal G}^{ij}_t(x)$.  
For the solutions of the problem (\ref{a.1})--(\ref{a.3}),
the following bound is true.
\begin{theorem}\label{t1}
Let $Y_0\in{\cal H}_{\alpha,+}$ and $\alpha>3/2$.  Then
\beqn \label{ubound}
\Vert U_0(t)Y_0\Vert_{-\alpha,+}\le
C\langle t\rangle^{-3/2}\Vert Y_0\Vert_{\alpha,+},\quad t\in\R.
\eeqn
\end{theorem}

 This theorem is proved in Appendix B.
\smallskip

To formulate the main result,  introduce the following notations.
\\
(i) Denote by ${\bf G}^j_{1}(y,t)$, $j=0,1$, the following function
\beqn\label{2.26}
{\bf G}^j_{1}(y,t)\!\!&:=&\!\!\Big(
{\cal G}^{j0}_{t,+}(1,y),{\cal G}^{j1}_{t,+}(1,y)\Big)\nonumber\\
\!\!&=&\!\!
\Big({\cal G}_t^{j0}(1-y)-{\cal G}_t^{j0}(1+y),
{\cal G}_t^{j1}(1-y)-{\cal G}_t^{j1}(1+y)\Big),\,\,y\in\Z,\,\,\,t\in\R.
\eeqn
(ii) Let ${\bf G}^j(y)$, $j=0,1$, stand for the vector valued function
 defined as
\beqn\label{Phi-j}
{\bf G}^j(y)=\int_0^{+\infty} N(s) {\bf G}_1^j(y,-s)\,ds=
\int_0^{+\infty} N^{(j)}(s) {\bf G}_1^0(y,-s)\,ds,\,\,\,\,y\in\Z,\,\,\,\, j=0,1,
\eeqn
where the function $ N(s)$ is introduced in (\ref{3.20})--(\ref{N}).
\smallskip\\
(iii) Denote by $U'_0(t)$ the operator adjoint to $U_0(t)$:
\be\label{defU'}
\langle Y,U'_0(t)\Psi\rangle_+=\langle U_0(t)Y,\Psi\rangle_+,
\quad Y\in{\cal H}_{\alpha,+},\quad\Psi\in{\cal S}\equiv [S(\Z_+)]^2,\quad t\in\R,
\ee
where $S(\Z_+)$ denotes the class of rapidly decreasing sequences on $\Z_+=\{x\in\Z:x\ge0\}$.
Using the Green function ${\cal G}_{t,+}$, we rewrite $U'_0(t)\Psi$ in the form
$$
(U'_0(t)\Psi)^j(y)=\sum\limits_{i=0,1}\sum\limits_{x\ge0}
{\cal G}^{ij}_{t,+}(x,y)\Psi^i(x),
\quad  t\in\R,\quad y\in\Z_+,\quad j=0,1.
$$
In particular,
${\bf G}^0_{1}(y,t)=(U'_0(t) Y_0)(y)$ with $Y_0(x)=(\delta_{1x},0)$ (see (\ref{2.26})),
where $\delta_{1x}$ denotes the Kronekker symbol.
\smallskip\\
(iv) Denote by  ${\bf K}^{j}(x,y)$  $j=0,1$, $x\in\N$, $y\in\Z$, vector-valued functions of a form
\beqn\label{Pi-0}
{\bf K}^j(x,y)&=&
\int_0^{+\infty}K(x,s)\Big(U'_0(-s){\bf G}^j\Big)(y)\,ds
\nonumber\\
&=&\int_0^{+\infty}\int_0^{+\infty}
K(x,s) N^{(j)}(\tau){\bf G}^0_1(y,-s-\tau)\,ds\,d\tau,
\,\,\,x\in\N,\,\,\,y\in\Z,\eeqn
where  $K(x,s)$ is defined in (\ref{K(x,t)}).\smallskip\\
(v)
Define an operator $\Omega:{\cal H}_{\alpha,+}\to{\cal H}_{-\alpha,+}$, $\alpha>3/2$, by the rule
\beqn\label{Omega}
\Omega:Y\to
Y+\nu^2\Big(\langle Y(\cdot),{\bf K}^0(x,\cdot)\rangle_+,\langle Y(\cdot),{\bf K}^1(x,\cdot)\rangle_+\Big).
\eeqn
Here we put  ${\bf K}^j(x,y)|_{x=0}:={\bf G}^j(y)$, $y\in\Z$.
The properties of the functions ${\bf K}^j$ and the operator $\Omega$ are specified
in Remark~\ref{rem6.6}.
The main result of the paper is the following theorem.
\begin{theorem}\label{thC}
 Let $Y_0\in{\cal H}_{\alpha,+}$, $\alpha>3/2$, and condition {\bf C} or ${\bf C}_0$ hold. Then
 the following assertions are fulfilled.\\
(i) $U(t)Y_0=\Omega(U_0(t)Y_0)+r(t)$,
where
$\Vert r(t)\Vert_{-\alpha,+}\le C\langle t\rangle^{-\beta/2}\Vert Y_0\Vert_{\alpha,+}$,
 $\beta=3$ if {\bf C} holds and $\beta=1$
if ${\bf C}_0$ holds,
$\Omega$ is a bounded operator defined by (\ref{Omega}).\\
(ii)
The solution of the problem (\ref{1.1})--(\ref{1.3}) obeys the bound (\ref{0.3}).
\end{theorem}

This theorem is proved in Section~\ref{sec5}.
The behavior of the solutions with the initial data  from the space ${\cal H}_{0,+}$ is discussed in Remark~\ref{rem4}.
If conditions {\bf C} and ${\bf C}_0$ are not fulfilled, then the bound (\ref{0.3}) for {\em any} initial data
from  ${\cal H}_{\alpha,+}$
is incorrect, see Remark~\ref{rem4.6}.

\setcounter{equation}{0}
\section{Fourier--Laplace transform} \label{sec3}

In this section,  we study the properties of the solutions $q(x,t)$
to the problem (\ref{b.1})--(\ref{b.4})
using the Fourier--Laplace transform.
\begin{definition}
Let $|q(t)|\le Ce^{B t}$.
The Fourier--Laplace transform of $q(t)$ is given by the formula
\be\label{La-F}
\tilde q(\omega)=\int_{0}^{+\infty}
e^{i\omega t}q(t)\,dt,\quad \Im\omega>B.
\ee
\end{definition}

The Gronwall inequality
implies standard a priori estimate for the solutions $q(x,t)$, $x\ge1$.
In particular,
there exist constants $A,B<\infty$ such that
$$
\sum\limits_{x\in\N}(|q(x,t)|^2+|\dot q(x,t)|^2)\le C e^{Bt}
\quad\mbox{ as }\,\, t\to+\infty.
$$
Hence, the Fourier--Laplace  transform of the solutions $q(x,t)$ to the problem (\ref{b.1}), (\ref{b.3})
with respect to $t$-variable,
$q(x,t)\to\tilde q(x,\omega)$, exists at least
for $\Im \omega>B$ and satisfies the following equation
\beqn\label{b.1'}
(-\nu^2\Delta_L+m^2-\omega^2)\tilde q(x,\omega)=0,
\quad x\in\N,\quad \Im\omega>B.
\eeqn
We construct the solution of (\ref{b.1'}).
We first note that the Fourier transform of the  operator $-\nu^2\Delta_L+m^2$
is the operator of multiplication by the function
$\phi^2(\theta)=\nu^2(2-2\cos\theta)+m^2$. Thus, $-\nu^2\Delta_L+m^2$ is a self-adjoint operator
and its spectrum is absolutely continuous and coincides with the range of $\phi^2(\theta)$,
i.e., with the segment $[m^2,m^2+4\nu^2]$.

\begin{lemma}\label{theta} (see Lemma 2.1 in  \cite{KKK})
Denote $\Lambda:=[-\sqrt{4\nu^2+ m^2},-m]\cup[m,\sqrt{4\nu^2+m^2}]$.
For given $\omega\in \mathbb{C}\setminus \Lambda$,
the equation
\be\label{32}
\nu^2(2-2\cos\theta)=\omega^2-m^2
\ee
has the unique solution  $\theta(\omega)$
in the domain $\{\theta\in\mathbb{C}:\,\Im\theta>0,\,
-\pi<\Re\,\theta\le\pi\}$.
Moreover, $\theta(\omega)$ is an analytic function in $\mathbb{C}\setminus \Lambda$.
\end{lemma}

Since we seek the solution  $q(\cdot,t)\in \ell^2_{\alpha,+}$
with some $\alpha$, $\tilde q(x,\omega)$ has a form
$$
\tilde q(x,\omega)=\tilde q(0,\omega) e^{i\theta(\omega)x},
\quad x\ge 0.
$$
Introduce a function $\tilde K(x,\omega)=e^{i\theta(\omega)x}$.
Applying the inverse Fourier--Laplace transform with respect to $\omega$-variable,
we write the solution $q(x,t)$ of the problem (\ref{b.1}), (\ref{b.3}) in the form
\be\label{qxt}
(q(x,t),\dot q(x,t))=
\int_{0}^t K(x,t-s)\big(q(0,s),\dot q(0,s)\big)\,ds,
\quad x\in\N,\quad t>0,
\ee
where
\be\label{K(x,t)}
K(x,t)=\frac1{2\pi}\int\limits_{-\infty+i\mu}^{+\infty+i\mu}
e^{-i\omega t}\tilde K(x,\omega)\,d\omega,
\quad \tilde K(x,\omega)=e^{i\theta(\omega)x},\quad x\in\N,
\quad t>0,
\ee
with some $\mu>0$.
The following theorem was proved in \cite{jmp-2017}.
\begin{theorem}\label{l2.15}
For any $\alpha<-3/2$, the following bound holds,
 \be\label{boundK}
 \sum\limits_{x\in\N}\langle x\rangle^{2\alpha}|K(x,t)|^2\le C(1+t)^{-3}
 \quad for \quad t>0.
 \ee
In particular,
\be\label{3.7'}
|K(1,t)|\le C(1+t)^{-3/2},\quad t>0.
\ee
\end{theorem}

To estimate $q(0,t)$, we use (\ref{qxt}) and rewrite Eqn~(\ref{b.2}) in the form
\be\label{7.1}
\ddot q(0,t)=-(\kappa+\nu^2+m^2)q(0,t)-\gamma \dot q(0,t)+\nu^2\int_0^t
K(1,t-s)q(0,s)\,ds+\nu^2 z(1,t),\quad t>0.
\ee
At first, we study the solutions of
  the corresponding homogeneous equation
\be\label{qt}
\ddot q(0,t)=-(\kappa+\nu^2+m^2)q(0,t)-\gamma \dot q(0,t)+\nu^2\int_0^t
K(1,t-s)q(0,s)\,ds,\quad t>0,
\ee
with the initial data
 \be\label{init}
q(0,t)|_{t=0}=u_0(0)=: q_0,\quad \dot q(0,t)|_{t=0}=v_0(0)=: p_0.
\ee
Applying the Fourier--Laplace transform
to the solutions $q(0,t)$ of (\ref{qt}), we obtain
\begin{equation}\label{3.20}
\tilde q(0,\omega)
=\tilde  N(\omega)\left(-i\omega q_0+q_0\gamma +p_0\right) \quad \mbox{for }\,\,\Im\omega>B,
\end{equation}
where, by definition,
$ \tilde  N(\omega) :=[\tilde D(\omega)]^{-1}$ and
\begin{equation}\label{3.21}
\tilde D(\omega):=-\omega^2+\kappa+\nu^2+m^2-i\omega\gamma
-\nu^2\tilde K(1,\omega),\quad
\tilde K(1,\omega)=e^{i\theta(\omega)},\quad\omega\in\mathbb{C}.
\end{equation}
The properties of the functions $\tilde D(\omega)$ and $\tilde N(\omega)$
are studied in Appendix A.
In particular, we prove that $\tilde N(\omega)$ is an analytic function in the upper half-space.
Denote
\be\label{N}
N(t)=\frac1{2\pi}\int\limits_{-\infty+i\mu}
^{+\infty+i\mu}e^{-i\omega t} \tilde N(\omega)\,d\omega,\quad t\ge0,
\quad \mbox{with some }\,\mu>0.
\ee
The following theorem is proved in Appendix~A.
\begin{theorem}\label{l3.1}
let condition {\bf C} or ${\bf C}_0$ hold.  Then
\be\label{NN}
|N^{(k)}(t)|\le C(1+t)^{-\beta/2},\quad t\ge0,\quad k=0,1,2,
\ee
where $\beta=3$ if {\bf C} holds and $\beta=1$
if  ${\bf C}_0$ holds.
\end{theorem}
\begin{cor} \label{cor3}
Denote by $S(t)$ a solving operator of the Cauchy problem
(\ref{qt}), (\ref{init}). Then the variation constants formula
gives the following representation
for the solution of the problem (\ref{7.1}), (\ref{init}):
$$
 \left(\ba{c}q(0,t)\\ \dot q(0,t)\ea\right)= S(t)
 \left(\ba{c}q_0\\ p_0\ea\right)+\int_0^tS(\tau)
\left(\ba{c}0\\ \nu^2z(1,t-\tau)\ea\right)\,d\tau,\quad t>0.
$$
Evidently, $S(0)=I$, and
the matrix $S(t)$ has a form
$\left(\ba{cc}\dot N(t)+\gamma N(t)& N(t)\\
\ddot N(t)+\gamma \dot N(t)&\dot N(t)\ea\right)$.
Moreover,   $|S(t)|\le C(1+t)^{-\beta/2}$, by Theorem~\ref{l3.1}.
\end{cor}

\setcounter{equation}{0}
\section{Asymptotic behavior of $Y(t)$ as $t\to\infty$}  \label{sec5}

Set
$q^{(0)}(x,t)=q(x,t)$, $q^{(1)}(x,t)=\dot q(x,t)$, $x\in\Z_+$.
\begin{pro} \label{pro-q}
Let $Y_0\in{\cal H}_{\alpha,+}$, $\alpha>3/2$, condition {\bf C} or ${\bf C}_0$ hold,
 and  $q(0,t)$ be a solution of the problem (\ref{7.1}), (\ref{init}).
 Then
\beqn\label{5.4}
q^{(j)}(0,t)= \nu^2\langle U_0(t) Y_0,{\bf G}^j\rangle_+ +r_j(t),\quad t>0,\quad
 |r_j(t)|\le C\langle t\rangle^{-\beta/2}\Vert Y_0\Vert_{\alpha,+},\,\,\,j=0,1,
\eeqn
where  the functions ${\bf G}^j$ are defined in (\ref{Phi-j}),
the number $\beta$ is introduced in Theorem~\ref{thC}.
\end{pro}
{\bf Proof}\,  Corollary \ref{cor3} and the bound (\ref{NN}) imply that
$$
q^{(j)}(0,t)
=\nu^2\int_0^t N^{(j)}(\tau)z(1,t-\tau)\,d\tau
+ O((1+t)^{-\beta/2}),\quad t>0,\quad j=0,1.
$$
Moreover,  the bounds (\ref{ubound}) and (\ref{NN}) give
$$
\Big|\int_t^{+\infty}\!\!\! N^{(j)}(\tau) z(1,t-\tau)d\tau\Big|
 \le C\int_t^{+\infty}\!\!\! \langle\tau\rangle^{-\beta/2} \langle t-\tau\rangle^{-3/2}d\tau
\le C\langle t\rangle^{-\beta/2}\Vert Y_0\Vert_{\alpha,+}.
$$
This implies the representation (\ref{5.4}), since by (\ref{2.26}) and (\ref{sol}), we have
$$
 z(1,t-\tau)=\langle U_0(t)Y_0(\cdot), {\bf G}_1^0(\cdot,-\tau)\rangle_+.
 \bo
$$
\begin{remark}\label{rem2.10}
{\rm Now we list the properties of the functions
${\bf G}^j_{1}(y,t)$ and ${\bf G}^j(y)$, $j=0,1$.\\
 (i) By (\ref{3.2}) and (\ref{hatcalG}),
the function ${\bf G}^j_{1}(y,t)$ is odd w.r.t. $y\in\Z$.
Then the function ${\bf G}^j$ is also odd.
Formulas (\ref{hatcalG}) and the Parseval identity give
\be\label{2.28}
\Vert{\bf G}^0_{1}(\cdot,t)\Vert^2_0=C
\int_{-\pi}^\pi
\Big( \cos^2(\phi(\theta) t)
+\frac{\sin^2(\phi(\theta)t)}{\phi^2(\theta)}\Big)\sin^2(\theta)\,d\theta\le C<\infty.
\ee
Here $\Vert\cdot\Vert_0$ denotes norm in $\ell^2\times\ell^2$.
\smallskip

(ii)
Let condition {\bf C} or ${\bf C}_0$ hold.
Since  ${\bf G}^0_1(y,t)=U'_0(t)(\delta_{1x},0)$, then
for any $\alpha>3/2$, имеем
\be\label{4.3''}
\Vert U'_0(t){\bf G}^j\Vert_{-\alpha,+}\le
\int_0^{+\infty}|N^{(j)}(s)|\Vert U'_0(t-s)(\delta_{1x},0)\Vert_{-\alpha,+}\,ds
\le C\langle t\rangle^{-\beta/2},
\ee
due to the bound (\ref{ubound}), because the action of the group $U'_0(t)$
coincides with action of the group $U_0(t)$, up to order of the components.
Therefore, for $\alpha>3/2$,
  \be\label{6.6}
  |\langle U_0(t)Y_0,{\bf G}^j\rangle_+|\le
  \Vert Y_0\Vert_{\alpha,+} \Vert U'_0(t){\bf G}^j\Vert_{-\alpha,+}
\le  C\langle t\rangle^{-\beta/2}\Vert Y_0\Vert_{\alpha,+}.
  \ee

(iii) Let condition {\bf C} hold.
Since $U'_0(t){\bf G}_1^0(y,-s)={\bf G}_1^0(y,t-s)$,
then  the bounds (\ref{NN}) and (\ref{2.28}) yield
\be\label{6.66}
 \sup_{t\in\R}\Vert  U'_0(t){\bf G}^j(\cdot)\Vert_0\le
 \sup_{t\in\R} \int_0^{+\infty}|N^{(j)}(s)|\Vert{\bf G}_1^0(\cdot,t-s)\Vert_0\,ds
 \le C\int_0^{+\infty}|N^{(j)}(s)|\,ds<\infty.
 \ee
}\end{remark}

\begin{lemma} \label{pro-qx}
Let $Y_0\in{\cal H}_{\alpha,+}$, $\alpha>3/2$, and condition {\bf C} or ${\bf C}_0$ hold.
 Then the
 solution $q(x,t)$ of the problem (\ref{b.1})--(\ref{b.4}) with $x\ge1$,
admits the following representation
\be\label{qj}
q^{(j)}(x,t)
= \nu^2\langle U_0(t) Y_0,{\bf K}^j(x,\cdot)\rangle_+ +r_j(x,t),
\quad j=0,1,\quad t>0,
\ee
where ${\bf K}^j$ is introduced in (\ref{Pi-0}),
$\Vert r_j(\cdot,t)\Vert_{-\alpha,+}\le C\langle t\rangle^{-\beta/2}\Vert Y_0\Vert_{\alpha,+}$.
Here, by definition, $\Vert r\Vert_{-\alpha,+}^2:=\sum_{x\in\N} \langle x\rangle^{-2\alpha}|r(x)|^2$.
\end{lemma}
{\bf Proof}\,
At first, by (\ref{qxt}) and (\ref{5.4}), we have
\beqn\label{6.1}
q^{(j)}(x,t)=\nu^2
\int_{0}^t K(x,t-s)\langle U_0(s)Y_0,{\bf G}^j(\cdot)\rangle_+\,ds
+r'_j(x,t),\quad x\in\N,
\eeqn
where $\Vert r'_j(\cdot,t)\Vert_{-\alpha,+}\le C\langle t\rangle^{-\beta/2}$.
Indeed,  (\ref{5.4}) and (\ref{boundK}) give
$$
\ba{rcl}
\Vert r'_j(\cdot,t)\Vert_{-\alpha,+}&=&
\ds\Big\Vert \int_{0}^t K(\cdot,t-s) r_j(s)\,ds\Big\Vert_{-\alpha,+}\le
\int_0^t \Vert K(\cdot,t-s)\Vert_{-\alpha,+}|r_j(s)|\,ds
\\
 &\le&\ds C\int_0^t (1+t-s)^{-3/2}(1+s)^{-\beta/2}\,ds\le C_1\langle t\rangle^{-\beta/2}.
\ea
$$
Second, the first term in the r.h.s. of (\ref{6.1}) has a form (see (\ref{Pi-0}))
\be\label{4.14}
\nu^2\int_{0}^t K(x,s)\langle U_0(t-s)Y_0,{\bf G}^j(\cdot)\rangle_+\,ds=
\nu^2\langle U_0(t)Y_0,{\bf K}^j(x,\cdot)\rangle_++r''_j(x,t),
\ee
where, by definition,
$ r''_j(x,t):=-\nu^2\ds\int_t^{+\infty}\!K(x,s)
\langle U_0(t-s)Y_0,{\bf G}^j\rangle_+\,ds$.
The bounds (\ref{boundK}) and (\ref{6.6}) yield
\beqn\label{6.5}
\Vert r''_j(\cdot,t)\Vert_{-\alpha,+}\le
\nu^2\int_t^{+\infty}\Vert K(\cdot,s)\Vert_{-\alpha,+}
\left|\langle U_0(t-s)Y_0,{\bf G}^j\rangle_+\right|\,ds
\le C\langle t\rangle^{-\beta/2}\Vert Y_0\Vert_{\alpha,+}.
\eeqn
Hence, the bounds (\ref{6.1})--(\ref{6.5}) imply (\ref{qj})
with $r_j(x,t)=r'_j(x,t)+r''_j(x,t)$.
\bo
\begin{remark}\label{rem6.6}
{\rm
(i) Set $\tilde K(0,\omega):=1$.
Then, formally, $K(0,t)=\delta_{0t}$.
Hence, we can put   ${\bf K}^j(0,y)={\bf G}^j(y)$, $y\in\Z$.
Then, the representation (\ref{5.4}) follows from (\ref{qj}).
\smallskip

(ii) By Remark \ref{rem2.10} and definition (\ref{Pi-0}),
the function ${\bf K}^j(x,y)$ is odd w.r.t. $y\in\Z$.
Furthermore, formulas (\ref{Pi-0}), (\ref{boundK}) and (\ref{4.3''}) imply that for $\alpha>3/2$,
$$
\Big\Vert\Vert U'_0(t){\bf K}^j(x,\cdot)\Vert_{-\alpha,+}\Big\Vert_{-\alpha,+}
\le
\int_0^{+\infty}\Vert K(x,s)\Vert_{-\alpha,+}
\Vert U'_0(t-s){\bf G}^j\Vert_{-\alpha,+}ds<C\langle t\rangle^{-\beta/2}.
$$
Hence,
for $\alpha>3/2$,
\beqn\label{4.11}
\Vert \langle U_0(t)Y_0, {\bf K}^j(x,\cdot)\rangle_+\Vert_{-\alpha,+}
=\Vert \langle Y_0(\cdot),U'_0(t) {\bf K}^j(x,\cdot)\rangle_+\Vert_{-\alpha,+}
\le
C\langle t\rangle^{-\beta/2}\Vert Y_0\Vert_{\alpha,+},\,\,\,t\in\R.\,
\eeqn
In particular, the operator $\Omega$ introduced in (\ref{Omega}) is bounded.
\smallskip

(iii)
If condition {\bf C} holds, then
$\Vert{\bf K}^j(x ,\cdot)\Vert_0\in{\cal H}_{-\alpha,+}$ with $\alpha>3/2$, since
$$
\Big\Vert\Vert{\bf K}^j(x ,\cdot)\Vert_0\Big\Vert_{-\alpha,+}\le
\int_0^{+\infty}\Vert K(x,s)\Vert_{-\alpha,+}
\Vert U'_0(-s){\bf G}^j\Vert_0\,ds<\infty
$$
by virtue of (\ref{Pi-0}), (\ref{boundK}) and (\ref{6.66}).
Therefore, (\ref{Omega}) implies that for $Y\in{\cal H}_{0,+}$,
$$
\Vert \Omega Y\Vert_{-\alpha,+}\le \Vert Y\Vert_{-\alpha,+}
+\nu^2\sum_{j=0,1}\Vert \langle Y(\cdot), {\bf K}^j(x,\cdot)\rangle_+\Vert_{-\alpha,+}
\le C\Vert Y\Vert_{0,+}.
$$
}\end{remark}
{\bf Proof of Theorem \ref{thC}}\,
The item (i) follows from the representations (\ref{2.1}), (\ref{5.4}) and (\ref{qj}).
Further, definition (\ref{Omega}), the bounds (\ref{ubound}), (\ref{6.6}) and (\ref{4.11}) give
\beqn\label{4.12}
\ba{rcl}
\Vert \Omega(U_0(t) Y_0)\Vert_{-\alpha,+}
&\le& \Vert U_0(t)Y_0\Vert_{-\alpha,+}+\nu^2\sum_{j=0,1}
\Vert \langle U_0(t)Y_0, {\bf K}^j(x,\cdot)\rangle_+\Vert_{-\alpha,+}
\\
&\le& C\langle t\rangle^{-\beta/2}\Vert Y_0\Vert_{\alpha,+}, \quad \alpha>3/2.
\ea\eeqn
 Thus, the bound (\ref{0.3}) follows
from the part (i) of Theorem \ref{thC} and the bound (\ref{4.12}).
\bo
\begin{remark}\label{rem4}
{\rm Let condition {\bf C} hold.
From the proofs of Lemmas \ref{pro-q} and \ref{pro-qx} we see that
the remainders $r_j(t)$ and $r_j(x,t)$ in decompositions (\ref{5.4}) and (\ref{qj})
are estimated by $z(1,t)$,
$$
|r_j(t)|\le C_1|Y_0(0)|\langle t\rangle^{-3/2} +C_2\int_t^{+\infty}\langle \tau\rangle^{-3/2}|z(1,t-\tau)|\,d\tau.
$$
$$
\ba{rcl}\Vert r_j(\cdot,t)\Vert_{-\alpha,+}&\le&
\ds C_1|Y_0(0)|\langle t\rangle^{-3/2} +C_2\int_0^t\langle t-s\rangle^{-3/2}\,ds
\int_s^{+\infty}\langle \tau\rangle^{-3/2}|z(1,s-\tau)|\,d\tau\\
&& +
\ds C_3\int_t^{+\infty}\langle s\rangle^{-3/2}ds\int_0^{+\infty}\langle \tau\rangle^{-3/2}|z(1,t-s-\tau)|\,d\tau.
\ea
$$
Hence, if $\sup_{t\in\R}|z(1,t)|=:M_0<\infty$, then
$U(t)Y_0=\Omega(U_0(t)Y_0)+r(x,t)$,
where
$$
\Vert r(\cdot,t)\Vert_{-\alpha,+}\le C_1|Y_0(0)|\langle t\rangle^{-3/2}+
 C M_0\langle t\rangle^{-1/2}\le C\langle t\rangle^{-1/2}.
$$
For instance, if initial data $Y_0(x)$ are such that
 $\hat u_{\rm odd}(\theta),\hat v_{\rm odd}(\theta)/\phi(\theta)\in L^1(\T)$, where
 $Y_{\rm odd}(x)=(u_{\rm odd}(x),v_{\rm odd}(x))$ is defined in (\ref{odd}),
then $|z(1,t)|\le C<\infty$.
In particular, this is true if $m\not=0$ and $Y_0\in {\cal H}_{0,+}$.
}
\end{remark}
\begin{remark}\label{rem4.6}
{\rm If conditions {\bf C} and ${\bf C}_0$ are not fulfilled, then the bound (\ref{0.3})
for any initial data $Y_0\in{\cal H}_{\alpha,+}$
is incorrect. Indeed, if $m=\kappa=0$, then
$\tilde N(\omega)$ has a simple pole at zero,
and any constant   is a solution of the system (\ref{1.1})--(\ref{1.2}).
If $\gamma=0$ and $\kappa>2\nu^2$, then there exists a number
$\omega_0>\sqrt{4\nu^2+m^2}$ such that $\tilde D(\omega_0)=0$,
and $\tilde N(\omega)$ has simple poles at the points
   $\omega=\pm\omega_0$, see Remark \ref{corB.8} below.
   Therefore, a function of the form $u(x,t)=e^{i\theta(\omega_0)x}\sin(\omega_0 t)$
 is the solution of the system,  where
 $\theta(\omega)$ is the solution of (\ref{32}), $\Re\theta(\omega_0)=\pi$,
 $\Im\theta(\omega_0)>0$.
 If one of the following two conditions holds:
  (1) $m=0$, $\kappa=2(\nu^2-\gamma^2)$ and $\gamma\in(0,\nu)$, or
 (2) $m\not=0$, $\kappa=\nu^2-\gamma^2\pm\sqrt{(\nu^2-\gamma^2)^2-m^2\gamma^2}$
  and $\gamma\in\left(0,\left(\sqrt{m^2+4\nu^2}-m\right)/2\right]$, then
   there exist points $\omega_*\in\Lambda\setminus\Lambda_0$ such that
  $\tilde D(\omega_*-i0)=0$ (see item (iv) of Lemma~\ref{l8.2}).
We denote  $\theta_+:=\lim_{\ve\to+0}\theta(\omega_*+i\ve)$, $\theta_+\in\R$.
  Then the function of the form $u(x,t)=\sin(\theta_+x+\omega_* t)$
  is a solution of the system (\ref{1.1})--(\ref{1.2}).
 }
\end{remark}

\appendix
\setcounter{section}{1}
\setcounter{equation}{0}
\setcounter{theorem}{0}
\section*{\large\bf Appendix A: Properties of
$\tilde D(\omega)$ and $\tilde N(\omega)$ for $\omega\in\mathbb{C}$}
%

Let $\Lambda:=[-\sqrt{4\nu^2+m^2},-m]\cup[m,\sqrt{4\nu^2+m^2}]$.
 $\Lambda_0=\{\pm m, \pm \sqrt{4\nu^2+m^2}\}$ denotes the set of the ``spectral edges''.
We first list the properties  of the function $e^{i\theta(\omega)}$
for  $\omega\in \mathbb{C}\setminus \Lambda$,
$\omega\in \Lambda\setminus\Lambda_0$, and $\omega\in \Lambda_0$.

 Let $\omega\in \mathbb{C}\setminus \Lambda$. Then $\Im\theta(\omega)>0$
and  $e^{i\theta(\omega)}$ is an analytic function.
Moreover, by (\ref{32}) and the condition $\Im\theta(\omega)>0$, we have
\be\label{7.0}
\left| e^{i\theta(\omega)}\right|\le C|\omega|^{-2}\quad \mbox{as }\,\,|\omega|\to\infty.
\ee
For $\omega\in\Lambda\setminus\Lambda_0$,
put  $\theta(\omega\pm i0)=\lim\limits_{\ve\to+0}\theta(\omega\pm i\ve)$.
Since $\overline{\theta(\omega)}=-\theta(\bar \omega)$
for $\omega\in \mathbb{C}\setminus \Lambda$, then
$e^{i\theta(\omega-i0)}=\overline{e^{i\theta(\omega+i0)}}$
for $\omega\in\Lambda\setminus \Lambda_0$.

We study the behavior of $e^{i\theta(\omega)}$ near
the  points in the set $\Lambda_0$.
From Eqn~(\ref{32}) we have
\beqn\label{a3}
e^{i\theta(\omega)}=\cos\theta(\omega)+i\sin\theta(\omega)
=1-\frac1{2\nu^2}(\omega^2-m^2)+\frac{i}{2\nu^2}
\sqrt{(\omega^2-m^2)(4\nu^2+m^2-\omega^2)}
\eeqn
for $\omega\in \mathbb{C}\setminus\Lambda$.
The Taylor expansion implies
\be\label{a8}
e^{i\theta(\omega)}= 1+\frac{i}{\nu}\sqrt{\omega^2-m^2}-
\frac1{2\nu^2}(\omega^2-m^2)-\frac{i}{8\nu^{3}}(\omega^2-m^2)^{3/2}+\dots\quad \mbox{as }\,\omega\to \pm m+i0,
\ee
where
$\omega\in\mathbb{C}_+:=\{\omega\in\mathbb{C}:\Im\omega>0\}$,
 $\Im\sqrt{\omega^2-m^2}>0$.
Here
${\rm sgn}(\Re\sqrt{\omega^2-m^2})={\rm sgn}(\Re\omega)$
for $\omega\in\mathbb{C}_+$.
 This choice of the branch of the complex root  $\sqrt{\omega^2-m^2}$ follows
 from the condition $\Im\theta(\omega)>0$.
Similarly,
\be\label{a9}
e^{i\theta(\omega)}=-1+\frac{i}{\nu}\sqrt{m^2+4\nu^2-\omega^2}+\frac1{2\nu^2}(m^2+4\nu^2-\omega^2)
-\frac{i}{8\nu^3}(m^2+4\nu^2-\omega^2)^{3/2}+\dots
\ee
as $\omega\to\pm\sqrt{m^2+4\nu^2}$, $\omega\in\mathbb{C}_+$.
Here the branch of the complex root $\sqrt{m^2+4\nu^2-\omega^2}$ is chosen so that
${\rm sgn}(\Re\sqrt{m^2+4\nu^2-\omega^2})={\rm sgn}(\Re\omega)$
that follows from the condition $\Im\theta(\omega)>0$.
If $m=0$, then (\ref{a3}) and the Taylor expansion imply
\be\label{a10}
e^{i\theta(\omega)}=
1+\frac{i\omega}{\nu}-\frac{\omega^2}{2\nu^2}-\frac{i\omega^3}{8\nu^3}+\dots
\quad \mbox{as }\,\,\omega\to0,
\ee
and $e^{i\theta(\omega)}= -1+i\sqrt{4\nu^2-\omega^2}/\nu+\dots$ as $\omega\to \pm 2\nu$,
$\omega\in\mathbb{C}_+$.

\begin{lemma}\label{l2.A}
(i) $\tilde N(\omega)$
 is meromorphic for $\omega\in \mathbb{C}\setminus \Lambda$.

(ii) $|\tilde N(\omega)|=O(|\omega|^{-2})$ as $|\omega|\to\infty$.

(iii) $\tilde D(\omega)\not=0$  for all  $\omega\in \mathbb{C}_+=\{\omega\in\mathbb{C}:\, \Im\,\omega>0\}$.

(iv) If $\gamma=0$, then
$\tilde D(\omega)\not=0$ for any $\omega\in \mathbb{C}_-=\{\omega\in\mathbb{C}:\, \Im\,\omega<0\}$.
\end{lemma}
{\bf Proof}\,
The first assertion of the lemma  follows from the analyticity of
$\tilde D(\omega)$  for $\omega\in \mathbb{C}\setminus \Lambda$.
The assertion (ii) follows from (\ref{3.21}) and (\ref{7.0}).
To prove the third assertion,
we assume opposite that  $\tilde D(\omega_0)=0$ for some $\omega_0\in \mathbb{C}_+$.
Hence, the function $u_*(x,t)=e^{i\theta(\omega_0)x}e^{-i\omega_0 t}$,
$x\ge0$, $t\ge0$, is a solution of the problem (\ref{1.1})--(\ref{1.2})
with the initial data $Y_*=e^{i\theta(\omega_0)x}(1,-i\omega_0)$.
Therefore, the Hamiltonian (\ref{H}) is
$$
H(u_*(\cdot,t),\dot u_*(\cdot,t))=e^{2t\,\Im\omega_0}H(Y_*)
\quad \mbox{for any } t>0,
\quad \mbox{where }\,\, H(Y_*)>0.
$$
Since $\Im\omega_0>0$ and $Y_*\in {\cal H}_{0,+}$,
  this exponential growth  contradicts the energy estimate (\ref{H-1}).
Hence, $\tilde D(\omega)\not=0$ for any $\omega\in \mathbb{C}_+$.

If $\gamma=0$, then
$\overline{\tilde D(\omega)}=\tilde D(\bar\omega)$,
since $\overline{\theta(\omega)}=-\theta(\bar\omega)$ for $\omega\in \mathbb{C}\setminus \Lambda$.
Therefore, item (iv) of the lemma follows from item (iii).
\bo

\begin{lemma}\label{l8.2}
Let the condition {\bf C} or ${\bf C}_0$ hold.
Then $\tilde D(\omega)\not=0$ for $\omega\in\R\setminus \Lambda$,
$\tilde D(\omega\pm i0)\not=0$ for $\omega\in\Lambda\setminus \Lambda_0$.
\end{lemma}
{\bf Proof}\,
(i)\, Let $\omega\in\R$ and $|\omega|>\sqrt{4\nu^2+m^2}$.
Then  $\Re\theta(\omega)=\pm\pi$.
Therefore,
$$
\tilde D(\omega)=-\omega^2+\kappa+\nu^2+m^2-i\omega\gamma+
\nu^2 e^{-\Im\theta(\omega)} \quad\mbox{with }\,\,\Im\theta(\omega)>0.
$$
 Hence, $\Im\tilde D(\omega)\not=0$ iff $\gamma\not=0$.
 On the other hand, $\Re\tilde D(\omega)=\kappa-2\nu^2$ for $\omega=\pm\sqrt{4\nu^2+m^2}$,
and $\Re\tilde D(\omega_1)<\Re\tilde D(\omega_2)$ if
$|\omega_1|>|\omega_2|\ge\sqrt{4\nu^2+m^2}$.
In particular, $\Re\tilde D(\omega)\to-\infty$ as $|\omega|\to\infty$.
Hence, for $|\omega|>\sqrt{4\nu^2+m^2}$, $\Re\tilde D(\omega)\not=0$ iff $\kappa\le 2\nu^2$.
Therefore, for such values of $\omega$,
$\tilde D(\omega)\not=0$ iff either $\gamma\not=0$ or $\gamma=0$ and $\kappa\le 2\nu^2$.
\smallskip

(ii)\, Let $m\not=0$ and  $\omega\in(-m,m)$.
Then, $\Re\theta(\omega)=0$.
Hence,
$$
\Re\tilde D(\omega)=-\omega^2+\kappa+\nu^2+m^2-\nu^2 e^{i\theta(\omega)}
>\kappa\quad\mbox{for }\,\,|\omega|<m,
 $$
 and $\Re\tilde D(\pm m)=\kappa$.
Therefore, $\tilde D(\omega)\not=0$  for any $|\omega|<m$,
since $\kappa\ge0$.
\smallskip

(iii)\, Let $\omega\in(-\sqrt{4\nu^2+m^2},-m)\cup(m,\sqrt{4\nu^2+m^2})$.
Then
$\Re\theta(\omega+i0)\in(-\pi,0)\cup(0,\pi)$ and $\Im\theta(\omega+i0)=0$.
Moreover, ${\rm sign}(\sin\theta(\omega+i0))={\rm sign}\,\omega$.
Hence, for $m\not=0$,
$$
\ba{lll}\Im\tilde D(\omega+i0)&=&-\omega\gamma -\nu^2\sin\theta(\omega+i0)
\\
&=&-{\rm sign}(\omega)\left(|\omega|\gamma+\sqrt{\omega^2-m^2}
\sqrt{\nu^2-(\omega^2-m^2)/4}\right)\not=0.
\ea
$$
If $m=0$, then
$\tilde D(\omega+i0)=\kappa-\omega^2/2-i\omega\left(\gamma+\sqrt{\nu^2-\omega^2/4}\right)\not=0$
for any $\kappa,\gamma\ge0$.
\smallskip

(iv)\,
Since
$\tilde D(\omega-i0)=\overline{\tilde D(\omega+i0)}-2 i\omega\gamma$
for $\omega\in\Lambda\setminus\Lambda_0$, then
\beqn
\tilde D(\omega\!-\!i0)&=&
-\omega^2+\kappa+\nu^2+m^2-
\nu^2 \cos\theta(\omega+i0)+i\nu^2\sin\theta(\omega+i0)-i\omega\gamma\nonumber\\
&=&
\kappa-(\omega^2-m^2)/2+i\left({\rm sign}(\omega)\frac12\sqrt{\omega^2-m^2}
\sqrt{4\nu^2+m^2-\omega^2}-\omega\gamma\right)\nonumber
\eeqn
for $\omega\in\Lambda\setminus\Lambda_0$. Hence,
$\tilde D(\omega-i0)=0$ for $\omega\in\Lambda\setminus\Lambda_0$ iff
\be\label{A29}
\kappa=(\omega^2\!-\!m^2)/2\quad\mbox{and }\,\,
\sqrt{\omega^2-m^2}
\sqrt{4\nu^2+m^2-\omega^2}=2|\omega|\gamma,\quad \omega^2\in(m^2,m^2+4\nu^2).
\ee
Then, $\gamma\not=0$.
Put $P:=\omega^2-m^2$. Hence, $P$ is a solution of the following equation
\be\label{A30}
P^2+4P(\gamma^2-\nu^2)+4m^2\gamma^2=0,\quad P\in(0,4\nu^2).
\ee
If $m=0$, then Eqn (\ref{A30}) has a unique solution $P=4(\nu^2-\gamma^2)\in(0,4\nu^2)$
iff $\gamma<\nu$. Then, $\kappa=(\omega^2\!-\!m^2)/2=P/2=2(\nu^2-\gamma^2)$
by the first equation in (\ref{A29}).
Thus, if $m=0$, $\kappa=2(\nu^2-\gamma^2)$ and $\gamma\in(0,\nu)$, then
there exist two points
$\omega=\pm\omega_*=\pm2\sqrt{\nu^2-\gamma^2}\in\Lambda\setminus\Lambda_0$
such that $\tilde D(\omega_*-i0)=0$.

If $m\not=0$, then  (\ref{A30}) has a  solution iff
$
(\gamma^2-\nu^2)^2-m^2\gamma^2\ge0$ and $\gamma\in(0,\nu)$.
This is equivalent to the conditions
$\gamma^2+m\gamma-\nu^2\le0$ and $\gamma\in(0,\nu)$, that coincides with the inequality
$\gamma\in\left(0,\left(\sqrt{m^2+4\nu^2}-m\right)/2\right]$.
Therefore, if $m\not=0$ and $\gamma\in\left(0,\left(\sqrt{m^2+4\nu^2}-m\right)/2\right]$, then
Eqn (\ref{A30}) has solutions
$$
P=2(\nu^2-\gamma^2)\pm 2\sqrt{(\nu^2-\gamma^2)^2-m^2\gamma^2}\in(0, 4\nu^2).
$$
Hence, $\kappa=(\omega^2\!-\!m^2)/2=P/2=\nu^2-\gamma^2\pm\sqrt{(\nu^2-\gamma^2)^2-m^2\gamma^2}$.

Thus,
  there are points $\omega_*\in\Lambda\setminus\Lambda_0$, in which $\tilde D(\omega_*-i0)=0$, iff
 $\gamma\not=0$ and one of the following conditions is fulfilled:
 (1) $m=0$, $\kappa=2(\nu^2-\gamma^2)$ and $\gamma\in(0,\nu)$;
 (2) $m\not=0$, $\kappa=\nu^2-\gamma^2\pm\sqrt{(\nu^2-\gamma^2)^2-m^2\gamma^2}$
  and $\gamma\in\left(0,\left(\sqrt{m^2+4\nu^2}-m\right)/2\right]$.
   These values of $\kappa,m,\gamma$ are eliminated by the condition {\bf C}.
\bo
\begin{remark}\label{corB.8}
{\rm
If  condition {\bf C} holds, then $\tilde D(\omega)\not=0$ for $\omega\in\Lambda_0$,
because
$\tilde D(\pm\sqrt{4\nu^2+m^2})=\kappa-2\nu^2\mp i\gamma\sqrt{4\nu^2+m^2}$  
and $\tilde D(\pm m)=\kappa\mp i\gamma m$.  
If condition  {\bf C} is not satisfied, then there are points
 $\omega\in\R$, in which $\tilde D(\omega)=0$.
For example, $\tilde D(0)=0$  in the case $m=\kappa=0$.
If $\gamma=\kappa=0$, then $\tilde D(\pm m)=0$.
If $\gamma=0$ and $\kappa=2\nu^2$, then $\tilde D(\pm \sqrt{m^2+4\nu^2})=0$.
If $\gamma=0$ and $\kappa>2\nu^2$, then $\exists\,\omega_0>\sqrt{4\nu^2+m^2}$ such that
  $\tilde D(\pm \omega_0)=0$,
 and $\tilde D'(\omega_0)=-2\omega_0(\kappa-\nu^2)/(2\kappa+m^2-\omega_0^2)<0$.
}\end{remark}

Now we study the asymptotic behavior of $\tilde D(\omega)$ and $\tilde N(\omega)=(\tilde D(\omega))^{-1}$
near the points $\omega\in\Lambda_0$.
In the neighborhood of the points $\omega=\pm\sqrt{4\nu^2+m^2}$ we use the representation (\ref{a9})
and obtain
\beqn\label{A.9}
\tilde D(\omega)
=\kappa-2\nu^2\mp i\gamma\sqrt{4\nu^2+m^2}- i\nu(4\nu^2+m^2-\omega^2)^{1/2}+\dots
\eeqn
as $\omega\to\pm\sqrt{4\nu^2+m^2}$, $\omega\in\mathbb{C}_+$.
Therefore, if $\gamma\not=0$ or $\gamma=0$ and $\kappa\not=2\nu^2$, then
$$
\tilde N(\omega)=(\tilde D(\omega))^{-1}= C_1+i\,C_2\sqrt{4\nu^2+m^2-\omega^2}+\dots,
\quad\omega\to\pm\sqrt{4\nu^2+m^2},\quad \omega\in\mathbb{C}_+,
$$
where $C_1=(\kappa-2\nu^2\mp i\gamma\sqrt{4\nu^2+m^2})^{-1}$ and  $C_2=\nu C_1^2$.
If $\gamma=0$ and $\kappa=2\nu^2$, then
$$
(\tilde D(\omega))^{-1}= \frac{i}{\nu}(4\nu^2+m^2-\omega^2)^{-1/2}+\frac1{2\nu^2}
+\dots,\quad\omega\to\pm\sqrt{4\nu^2+m^2}.
$$

In the neighborhood of the points $\omega=\pm m$
we apply (\ref{a8}) (if $m\not=0$) and obtain
\be\label{B.44}
\tilde D(\omega)
= \kappa\mp i m\gamma-i\nu\sqrt{\omega^2-m^2}-i(\omega\mp m)\gamma
-\frac12(\omega^2-m^2)+\dots,\,\,\,
\omega\to\pm m,\,\,\, \omega\in\mathbb{C}_+.
\ee
In the case when $m=0$, (\ref{a10}) yields
\be\label{B.55}
\tilde D(\omega)
= \kappa-i\omega(\gamma+\nu)-\frac12\omega^2+\frac{i}{8\nu}\omega^3+..,\quad \omega\to0.
\ee
Suppose that
$m\gamma \not=0$  or
 $\kappa\not=0$.  Then, by virtue of  (\ref{B.44}) and (\ref{B.55}), we obtain
$$
\tilde N(\omega)=(\tilde D(\omega))^{-1}=\left\{
\ba{lll}
1/\kappa+i\,\omega(\gamma+\nu)/\kappa^2+\dots,&\omega\to0,& m=0,\\
 C_3+i\,C_4(\omega^2-m^2)^{1/2}+\dots,&\omega\to\pm m,& m\not=0,
\ea\right.\quad \omega\in\mathbb{C}_+,
$$
where $C_3=(\kappa\mp im\gamma)^{-1}$ and $C_4=\nu C_3^{2}$.
If $\gamma=\kappa=0$ and $m\not=0$, then
$$
\tilde N(\omega)=\frac{i}{\nu}(\omega^2-m^2)^{-1/2}-\frac1{2\nu^2}
+\dots,\quad\omega\to\pm m,\quad \omega\in\mathbb{C}_+.
$$
If $\kappa=m=0$, then 
$\tilde N(\omega)=i\omega^{-1}/(\gamma+\nu)-1/(2(\gamma+\nu)^{2})+\dots$
as $\omega\to0$.

Since $\tilde N(\omega)=(\overline{\tilde D(\bar\omega)}-2i\omega\gamma)^{-1}$
for $\omega\in\mathbb{C}_-$,
then the expansion for $\tilde N(\omega)$
as $\omega\to\omega_0$ ($\omega_0\in\Lambda_0$, $\omega\in\mathbb{C}_-$)  can be constructed
using  (\ref{A.9}) and (\ref{B.44}). In particular,
\be\label{A.11}
\tilde N(\omega+i0)-\tilde N(\omega-i0)=O(|\omega^2-\omega_0^2|^{j/2})\quad
\mbox{as }\,\,\omega\to\omega_0,\,\,\,\omega_0\in\Lambda_0,
\ee
where $j=1$  if the condition {\bf C} is satisfied, and $j=-1$ if the condition ${\bf C}_0$ is satisfied.
\medskip\\
{\bf Proof of Theorem \ref{l3.1}}\,
Using Lemma \ref{l2.A}, we vary the integration contour in the right hand side of (\ref{N}):
\be\label{b.10}
N(t)=-\frac1{2\pi}\int_{|\omega|=R}
e^{-i\omega t} \tilde N(\omega)\,d\omega,\quad t>0,
\ee
where $R$ is chosen enough large such that $ \tilde N(\omega)$ has no poles
in the region $\mathbb{C}_-\cap\{|\omega|\ge R\}$.
Note that if $\gamma=0$, then $\tilde N(\omega)$ has no poles
in $\mathbb{C}_-$ by Lemma \ref{l2.A} (iv).
Denote by $\sigma_j$ the poles of $\tilde N(\omega)$ in $\mathbb{C}_-$
(if they exist).
By Lemmas \ref{l2.A} and \ref{l8.2}, there exists a $\delta>0$ such that
$\tilde N(\omega)$ has no poles in the region $\Im\omega\in[-\delta,0)$.
Hence, we can rewrite $N(t)$ as
$$
N(t)=-i\sum\limits_{j=1}^K{\rm Res}_{\omega=\sigma_j}
\left[e^{-i\omega t} \tilde N(\omega)\right]
-\frac1{2\pi}\int_{\Lambda_{\ve}}
e^{-i\omega t} \tilde N(\omega)\,d\omega,\quad t>0,
$$
where $\varepsilon\in(0,\delta)$,
 the contour $\Lambda_{\ve}$ surrounds segments of $\Lambda$
and belongs to an $\ve$-neighborhood of $\Lambda$
($\Lambda_{\ve}$ is  oriented anticlockwise).
Passing to a limit as $\ve\to0$, we obtain
  \beqn
 N(t)&=&\frac1{2\pi}\int_{\Lambda}
 e^{-i\omega t}\left(\tilde N(\omega+i0)-\tilde N(\omega-i0)\right)\,d\omega
 +o(t^{-N})\nonumber\\
 &=&
  \sum\limits_{\pm}\sum\limits_{j=1}^2\frac{1}{2\pi}\int_{\Lambda}
 e^{-i\omega t} P_j^{\pm}(\omega)\,d\omega+o(t^{-N}),\quad t\to+\infty,
 \quad \mbox{with any }\, N>0.  \nonumber
\eeqn
Here $P_j^{\pm}(\omega):=\zeta_j^\pm(\omega)(\tilde N(\omega+i0)-\tilde N(\omega-i0))$,
 $j=1,2$, where $\zeta_j^\pm(\omega)$ are smooth functions such that
$\sum\limits_{\pm,j}\zeta_j^\pm(\omega)=1$, $\omega\in\R$,
$\supp\zeta_1^\pm\subset {\cal O}(\pm m)$,
$\supp\zeta_2^\pm\subset {\cal O}(\pm \sqrt{4\nu^2+m^2})$
(${\cal O}(a)$ denotes a neighborhood of the point $\omega=a$).
In the case $m=0$, instead of $\zeta_1^\pm$
($P_1^\pm$) we introduce  the function $\zeta_1$ (respectively, $P_1$)
  with $\supp\zeta_1\subset {\cal O}(0)$.
Then, (\ref{A.11}) implies the bound (\ref{NN}) with $k=0$.
Here we use the following estimate (with $j=\pm1$)
\be\label{b.11}
\Big|\int_{\R}\zeta(\omega) e^{-i\omega t}
(a^2-\omega^2)^{j/2}\,d\omega\Big|\le C(1+t)^{-1-j/2}\quad
\mbox{as }\,t\to+\infty,\,\,\,\mbox{$j$ is odd},
\ee
 where $\zeta(\omega)$ is a smooth function,
and $\zeta(\omega)=1$ for $|\omega-a|\le\delta$ with some $\delta>0$
(see, for example,  \cite[Lemma 2]{V74}).
The bound (\ref{NN}) with $k=1,2$ can be proved by a similar way.
\bo
\begin{remark}
{\rm If conditions ${\bf C}$ and ${\bf C}_0$
are not fulfilled, then $N(t)$ does not decay as $t\to\infty$.
For example, if $\kappa=m=0$,
 then
$\tilde N(\omega)$ has a simple pole at zero. Calculating the residue of
$\tilde N(\omega)$ at the point $\omega=0$, we obtain
$N(t)=(\gamma+\nu)^{-1}+O(t^{-3/2})$, $t\to\infty$.

If $\gamma=0$ and $\kappa>2\nu^2$, then there exists a number
$\omega_0>\sqrt{4\nu^2+m^2}$ such that  $\tilde N(\omega)$ has simple poles at the points
   $\omega=\pm\omega_0$.
Calculating the residue of
$e^{-i\omega t}\tilde N(\omega)$ at these points, we obtain
$N(t)\sim C\sin(\omega_0 t)+O(t^{-3/2})$ as $t\to\infty$.
}
\end{remark}
\newpage
\setcounter{section}{2}
\setcounter{equation}{0}
\setcounter{theorem}{0}
\section*{Appendix B: Proof of Theorem 2.4}

Consider the  mixed initial-boundary value problem  (\ref{a.1})--(\ref{a.3}).
Without loss of generality, we assume that $u_0(0)=v_0(0)=0$.
Write $Z(x,t)=\left(Z^0(x,t),Z^1(x,t)\right)\equiv(z(x,t),\dot z(x,t))$,
$Y_0(x)=(u_0(x),v_0(x))$.
The  solution of  problem (\ref{a.1})--(\ref{a.3})
can be represented as the restriction of the solution to the Cauchy problem
with odd initial data on the half-line,
\be\label{7.2}
Z^i(x,t)=\sum\limits_{y\in\Z}
{\cal G}^{ij}_{t}(x-y) Y^j_{\rm odd}(y),\quad
x\ge0,\quad i=0,1,
\ee
where ${\cal G}_t(x)$ is defined in  (\ref{3.2}) and (\ref{hatcalG}),
  and, by definition,
\beqn\label{odd}
Y_{\rm odd}(x)=Y_0(x)\quad\mbox{for }\,x>0,\quad Y_{\rm odd}(0)=0,\quad
Y_{\rm odd}(x)=-Y_0(-x)\quad\mbox{for }\, x<0.\,\,
\eeqn


To prove Theorem \ref{t1} we first consider the following Cauchy problem
 for the discrete Klein--Gordon equation in the whole line,
\begin{equation}\label{KG}
    \left\{\ba{l}
\ddot u(x,t) =(\nu^2\Delta_L-m^2)u(x,t),
 \quad t\in\mathbb{R},\quad x\in\mathbb{Z},\\
u(x,t)|_{t=0}=u_0(x),\quad
\dot u(x,t)|_{t=0}=v_0(x).
  \ea\right.
  \end{equation}
By  $\ell^2_{\alpha}\equiv\ell^2_{\alpha}(\Z)$, $\alpha\in\R$,
we denote the  Hilbert space of sequences with the norm
$\Vert u\Vert^2_{\alpha}
=\sum\limits_{x\in\Z}|u(x)|^2 \langle  x\rangle^{2\alpha}<\infty$.
Let ${\cal H}_{\alpha}:=\ell^2_{\alpha}\otimes\ell^2_{\alpha}$
be the Hilbert space of pairs
  $Y=(u,v)$ with the norm $\Vert Y\Vert^2_{\alpha}
=\sum\limits_{x\in\Z} \langle  x\rangle^{2\alpha}(|u(x)|^2+|v(x)|^2)<\infty$.

It is well-known (see for instance, \cite{D08}), that for any $Z_0\equiv(u_0,v_0)\in {\cal H}_{\alpha}$,
 there exists a
unique solution $W(t)Z_0\in C(\mathbb{R},{\cal H}_{\alpha})$ to the problem (\ref{KG}).
Moreover, there exist constants
 $C,\sigma=\sigma(\alpha)<\infty$ such that the following bound holds,
 \be\label{est-W-kg}
 \Vert W(t)Z_0\Vert_{\alpha}\le C\langle t\rangle^\sigma\Vert Z_0\Vert_{\alpha},
 \quad t\in\mathbb{R},\quad \alpha\in\mathbb{R}.
 \ee
\begin{lemma}\label{lemmaC.1}
 Let $Z_0\equiv(u_0,v_0)\in{\cal H}_\alpha$ with $\alpha>5/2$.
If $\hat Z_0(0)=\hat Z_0(\pi)=0$, then
\be\label{3.16}
\Vert W(t)Z_0\Vert_{-\alpha}\le C\langle t\rangle^{-3/2}\Vert Z_0\Vert_{\alpha},
\quad t\in\mathbb{R}.
\ee
Otherwise,
$\Vert W(t) Z_0\Vert_{-\alpha}\le C\langle t\rangle^{-1/2}\Vert Z_0\Vert_{\alpha}$,
$t\in\mathbb{R}$.
\end{lemma}

Below we outline the proof of this lemma.

By the bound (\ref{est-W-kg}),
 the Laplace--Fourier transform of the solution $u(x,t)$ with respect to $t$-variable
 exists at least for $\Im \omega>0$ and satisfies  equation
(\ref{b.1'}) for $x\in\mathbb{Z}$,  $\Im\omega>0$.
Let $u$ be a solution of the equation $(-\nu^2\Delta_L+m^2-\omega^2)u=f$
with $f\in\ell^2$. Define the resolvent operator $R_\omega$ as
 $u=R_\omega f=(-\nu^2\Delta_L+m^2-\omega^2)^{-1}f$.

Applying the inverse Fourier--Laplace transform with respect to $\omega$-variable,
we write the solution $u(x,t)$ of the problem (\ref{KG}) in the form
\beqn\label{2.23}
u(x,t)=\frac1{2\pi}\int_{\Im\omega=\mu}
e^{-i\omega t}R_\omega (v_0(x)-i\omega u_0(x))\,d\omega,
\quad x\in\mathbb{Z}, \quad t>0,\quad \mu>0.
\eeqn
To derive the asymptotic behavior of $u(x,t)$, we first study the properties of the operator
  $R_\omega$ for  $\omega\in\mathbb{C}$, see \cite{IV, ShV, KKK}.
To formulate them, we denote by $B(\alpha,\alpha')={\cal L}(\ell^2_\alpha,\ell^2_{\alpha'})$ the space
of bounded linear operators from $\ell^2_\alpha$ to $\ell^2_{-\alpha}$.
\smallskip\\
{\bf I}.
For $\omega\in\mathbb{C}\setminus \Lambda$,
 the resolvent $R_\omega$ is the integral operator with the kernel $R_\omega(x,y)$,
 $x,y\in\mathbb{Z}$, and by the Cauchy Residue Theorem, we have
\be\label{R}
R_\omega(x,y)=
\frac1{2\pi}\int_{\mathbb{T}}
\frac{e^{-i\theta(x-y)}}{\nu^2(2-2\cos\theta)+m^2-\omega^2}\,d\theta
=i\frac{e^{i\theta(\omega)|x-y|}}{2\nu^2\sin(\theta(\omega))},
\quad\omega\in\mathbb{C}\setminus \Lambda,
\ee
 where $\theta(\omega)$ is defined in Lemma \ref{theta}.
Therefore, for $\omega\in\mathbb{C}\setminus \Lambda$, the resolvent $R_\omega$ is an analytic
operator-valued function in the complex
$\omega$-plane with the cut along the intervals in $\Lambda$.
 Moreover, 
the sequence $\{e^{-i\theta(\omega)|x|}\}$, $x\in\mathbb{Z}$,
is exponentially decaying as $|x|\to\infty$.
 Hence for $\omega\in\mathbb{C}\setminus \Lambda$,
 $R_\omega$ is a bounded operator in $\ell^2(\mathbb{Z})$.
 \smallskip\\
{\bf II}.
Write $\theta(\omega\pm i0):=\lim\limits_{\ve\to+0}\theta(\omega\pm i\ve)$.
For $\omega\in\Lambda\setminus \Lambda_0$ and $x,y\in\mathbb{Z}$,
the following pointwise limit exists
$R_{\omega\pm i \ve} (x,y)\to R_{\omega\pm i0}(x,y)$
as $\ve\to+0$.
Moreover, $|\theta(\omega\pm i\ve)|\le C(\omega)$ and
$|\sin\theta(\omega\pm i\ve)|>0$ for $\omega\in\Lambda\setminus \Lambda_0$.
Hence, $|R_{\omega\pm i \ve} (x,y)|\le C(\omega)$ for $\omega\in\Lambda\setminus \Lambda_0$. Therefore,
for any $\alpha>1/2$ and $\omega\not\in \Lambda_0$, we have
$$
 \sum\limits_{x,y\in\mathbb{Z}}
  \left|R_{\omega\pm i\ve}(x,y)-R_{\omega\pm i0}(x,y)\right|^2
   \langle x\rangle^{-2\alpha} \langle y\rangle^{-2\alpha}
   \to0,\quad \ve\to+0,
$$
by the Lebesgue dominated convergence  theorem.
Thus,
for $\omega\in\Lambda\setminus \Lambda_0$,
the resolvent $R_{\omega\pm i\ve}$ converges to $R_{\omega\pm i0}$
($\ve\to+0$) as Hilbert--Schmidt operator in the space $B(\alpha,-\alpha)$,
 $\alpha>1/2$.
Moreover,  $\overline{\theta(\omega)}=-\theta(\bar\omega)$
for $\omega\in \mathbb{C}\setminus \Lambda$. Hence,
$R_{\omega-i0}(x,y)=\overline{R_{\omega+i0}(x,y)}$ for
$\omega\in\Lambda\setminus \Lambda_0$, $x,y\in\mathbb{Z}$.
\smallskip\\
{\bf III}. The operator
$R_{\omega\pm i0}$ diverges near points $\omega\in\Lambda_0$
because $\sin\theta(\omega+i0)$ vanishes in these points.
Using formula (\ref{R}) and decompositions (\ref{a8})--(\ref{a10}),
we obtain a formal Puiseux expansion
of $R_\omega$ as $\omega\to\omega_0$,  $\omega\in\mathbb{C}\setminus \Lambda$,
$\omega_0\in\Lambda_0$.
Indeed, for $\omega\to\pm m$ ($m\not=0$, $\omega\in\mathbb{C}_+$),
we have  
\be\label{B.12}
R_\omega(x,y)
=\frac{i}{2\nu}(\omega^2-m^2)^{-1/2}-\frac1{2\nu^2}|x-y|-\frac{i}{16\nu^3}(4|x-y|^2-1)(\omega^2-m^2)^{1/2}+\dots,
\ee
where
$\Im\sqrt{\omega^2-m^2}>0$.
In particular, if $m=0$, then
$$
R_\omega(x,y)=\frac{i}{2\nu \omega}-\frac1{2\nu^2}|x-y|-\frac{i\omega}{16\nu^3}(4|x-y|^2-1)+\dots,
\quad \omega\to0.
$$
For $\omega\to\pm \sqrt{4\nu^2+m^2}$, $\omega\in\mathbb{C}_+$,
\beqn\label{B.13}
\begin{array}{lll}
R_\omega(x,y)&=&(-1)^{|x-y|}\Big(
\ds\frac{i}{2\nu}(4\nu^2+m^2-\omega^2)^{-1/2}+\frac1{2\nu^2}|x-y|\\
&&-
\ds\frac{i}{16\nu^3}(4|x-y|^2-1)\sqrt{4\nu^2+m^2-\omega^2}+\dots
\Big).
\end{array}
\eeqn
Since
$\sum\limits_{x,y\in\mathbb{Z}}\langle x\rangle^{-2\alpha}|x-y|^{2p}\langle y\rangle^{-2\alpha}<\infty$
for $\alpha>\frac12+p$, with any $p=0,1,2,\dots$,
\be\label{B.14}
\Vert|x-y|^{p} f(y)\Vert_{-\alpha}\le C\Vert f\Vert_\alpha,\quad f\in\ell^2_\alpha,\quad \alpha>\frac12+p,
\quad p=0,1,2,\dots.
\ee
Applying these estimates to the terms in the expansions (\ref{B.12}) and (\ref{B.13}), we come to the following result.
\begin{lemma}\label{t3.3} (see \cite[Lemma 3.2]{KKK})
Let $f\in\ell^2_{\alpha}$, $\alpha>5/2$.
 Then for $\omega\to\pm m$, $\omega\in\mathbb{C}\setminus\Lambda$, we have
$$
(R_{\omega} f)(x)=\frac{i\hat f(0)}{2\nu\sqrt{\omega^2-m^2}}
-\frac1{2\nu^2}\sum\limits_{y\in\mathbb{Z}}|x-y| f(y)+\sqrt{\omega^2-m^2}\,\,r^1_\omega f,
$$
 and for $\omega\to\pm\sqrt{4\nu^2+m^2}$, $\omega\in\mathbb{C}\setminus\Lambda$,
$$
(R_{\omega} f)(x)=\frac{i(-1)^x\hat f(\pi)}{2\nu\sqrt{4\nu^2+m^2-\omega^2}}
+\frac1{2\nu^2}\sum\limits_{y\in\mathbb{Z}}(-1)^{|x-y|}|x-y| f(y)+\sqrt{4\nu^2+m^2-\omega^2}\,\,r^2_\omega f,
$$
where the remainder terms have the form
$r^j_\omega f=\sum_{k=0}^{2}b^j_k(\omega)\sum_{y\in\mathbb{Z}}|x-y|^k f(y)$,
 $b^1_k(\omega)=O(1)$ as $\omega\to\pm m$ and $b^2_k(\omega)=O(1)$ as $\omega\to\pm\sqrt{4\nu^2+m^2}$.
In particular,
$\Vert r^j_\omega f\Vert_{-\alpha}\le C\Vert f\Vert_{\alpha}$.
\end{lemma}

Now Lemma \ref{lemmaC.1} follows from the equality (\ref{2.23}) and
Lemma \ref{t3.3}, using arguments similar to the proof Theorem \ref{l3.1}
and technique of the paper \cite{KKK}.
\medskip

{\bf Proof of the bound (\ref{ubound})}.
Using the representation
(\ref{7.2}) and formula (\ref{2.23}), we rewrite the solution of the problem
 (\ref{a.1})--(\ref{a.3}) in the form
$$
z(x,t)=\frac1{2\pi}\int_{\Im\omega=\mu}
e^{-i\omega t}R_\omega f_{\rm odd}
\,d\omega,
\quad x\in\mathbb{Z_+}, \quad t>0,\quad \mu>0,
$$
where $f_{\rm odd}(x):=v_{\rm odd}(x)-i\omega\, u_{\rm odd}(x)$ (see (\ref{odd})).
Applying arguments similar to the proof
of Theorem~\ref{l3.1}, we obtain
 \beqn\label{B.9}
z(x,t)=\frac1{2\pi}\int_{\Lambda}
e^{-i\omega t}\Big(R_{\omega+i0}-R_{\omega+i0}\Big)f_{\rm odd}
\,d\omega=\frac1{\pi}\int_{\Lambda}
e^{-i\omega t} \,\Im \Big(R_{\omega+i0}f_{\rm odd}\Big)\,d\omega.
\eeqn
Let   $Y_0\in{\cal H}_{\alpha,+}$ with $\alpha>3/2$.
Then, $Y_{\rm odd}\in{\cal H}_{\alpha}$ and $\hat f_{\rm odd}(0)=\hat f_{\rm odd}(\pi)=0$.
We want to apply Lemma \ref{t3.3} to the function $f_{\rm odd}(x)$, but with $\alpha>3/2$ instead of
$\alpha>5/2$, using the oddness of $f_{\rm odd}(x)$.
Note that for $k=1,2$,
$$
\Big|\sum\limits_{y\in\Z}|x-y|^k f_{\rm odd}(y)\Big|\le
2|x|\sum\limits_{y\ge1}y |f_0(y)|,
$$
where $f_0(x):=v_0(x)-i\omega\, u_0(x)$, $x\in\Z_+$. Therefore,
  applying the Cauchy--Bunyakovskii inequality, we obtain for
$\alpha>3/2$,
\beqn
\ba{rcl}
\Vert r^j_\omega f_{\rm odd}\Vert_{-\alpha,+}
\!\!&\le&\!\! C\sum\limits_{k=0}^2\Big\Vert\sum\limits_{y\in\Z}|x-y|^kf_{\rm odd}(y)\Big\Vert_{-\alpha,+}
\le
C_1\sqrt{\sum\limits_{x\in\Z_+}\langle x\rangle^{-2\alpha}\,x^2\Big(\sum\limits_{y\in\Z_+}y |f_0(y)|\Big)^2}
\\
\!\!&\le&\!\! C_2\sum\limits_{y\in\Z_+}
\langle y\rangle^{-\alpha}|y|\cdot \langle y\rangle^{\alpha}|f_0(y)|
\le C\Vert f_0\Vert_{\alpha,+}.\nonumber
\ea\eeqn
Thus, in the neighborhood of the singular points $\omega_0\in\Lambda_0$ the following estimate holds
\beqn\label{7.11}
\Vert\Im R_{\omega+i0} f_{\rm odd}\Vert_{-\alpha,+}\le C
|\omega^2-\omega^2_0|^{1/2}\Vert f_0\Vert_{\alpha,+},\,\,\,\,\omega\to\omega_0,
\eeqn
where $\alpha>3/2$, $\omega_0\in\Lambda_0$, $\omega\in\R$.
Now the estimate (\ref{ubound})
follows from the equality (\ref{B.9}),
   estimate (\ref {7.11}) and Lemma 10.2 from \cite{JK},
which is a generalization of the estimate (\ref{b.11}).
\bo


\end{document}